\documentclass[prb,twocolumn, preprintnumbers,amsmath,amssymb, showpacs]{revtex4}
\usepackage{amsmath}
\usepackage{graphicx}
\usepackage{dcolumn}
\DeclareGraphicsExtensions{.png .jpg .pdf}

\begin{document}

\title{Energy shift and conduction-to-valence band transition mediated by a time dependent potential barrier in graphene}

\author{Andrey Chaves}\email{andrey@fisica.ufc.br}
\author{D. R. da Costa}\email{diego_rabelo@fisica.ufc.br}
\author{G. O. de Sousa} \email{gabrieloliveira@fisica.ufc.br}
\author{J. M. Pereira Jr.}
\author{G. A. Farias} 
\affiliation{Departamento de F\'isica, Universidade
Federal do Cear\'a, Caixa Postal 6030, Campus do Pici, 60455-900
Fortaleza, Cear\'a, Brazil}

\date{ \today }

\pacs{81.05.U-, 73.63.-b, 72.80.Vp, 03.65.Pm}

\begin{abstract}
We investigate the scattering of a wave packet describing low-energy electrons in graphene by a time-dependent finite step potential barrier. Our results demonstrate that, after Klein tunneling through the barrier, the electron acquires an extra energy which depends on the rate of change the barrier height in time. If such a rate is negative, the electron loses energy and ends up as a valence band state after leaving the barrier, which effectively behaves as a positively charged quasi-particle.

\end{abstract}

\maketitle

The phenomenon of tunneling has long been a textbook example of the contrasts between quantum and classical behaviors. The fact that a particle can, with a certain probability, be transmitted through a classically forbidden region is not only a manifestation of the dual wave-particle character of microscopic systems, but is also at the basis of the explanation of phenomena such as the alpha emission. In most discussions of tunneling through one-dimensional potential profiles, the barrier is assumed to be static and the usual solution for the transmission coefficients can be found by comparing the amplitudes of incoming and outgoing plane waves. Furthermore, textbook calculations are usually presented within the framework of non-relativistic quantum mechanics \cite{textbook}. 

Recently there has been a renewed interest in the study of tunneling in the relativistic regime. \cite{Tunneling1, Tunneling2, Tunneling3, Chaves2, Logemann} This was motivated by the production of graphene, \cite{AKGeim} a bi-dimensional layer of carbon atoms organized in a honeycomb lattice. One unusual aspect of the electronic band structure of that material is the fact that at the vicinity of the Fermi level the spectrum is gapless with a linear dispersion, such that the charge carriers can be described as ultra-relativistic massless fermions, albeit with an effective light speed given by the Fermi velocity of the material ($v_F \approx c/300$). \cite{Misha, CastroNetoReview} This feature has important consequences, such as the perfect transmission of normally incident electrons through p-n and p-n-p junctions, an effect known as Klein tunneling. \cite{Chaves2} Thus graphene, apart from being of interest for device applications, also allows the investigation of quantum relativistic effects by means of tabletop experiments.  

Quantum mechanical tunneling through one-dimensional oscillating or moving potential barriers has been previously investigated in the context of non-relativistic electrons described by  Schr\"odinger's equation. \cite{time1,time2,time3, time4, time5} These studies have shown that a wavepacket incident on an oscillating barrier develops multiple peaks which propagate with different velocities. This has been explained as resulting from an energy exchange between the incident particle and the oscillating potential. Due to the parabolic dispersion of the tunneling particles in these calculations, an energy shift of the wavepacket components will result on different group velocities. 

In this work we investigate the interaction of wavepackets corresponding to massless Fermion states propagating in graphene with time-dependent potential barriers. We show that the gapless and linear aspects of the dispersion in graphene lead to an outcome that is quite distinct from the non-relativistic case: the electron-hole symmetry in graphene implies that, within a single-particle description, depending on the initial energy and the rate of change of potential, the resulting energy shift can convert an incoming conduction band electron into an outgoing particle within the valence band, which behaves as a positively charged quasi-particle. In addition, the fact that at low energies the group velocity of the charge carriers is independent of the energy, together with the perfect transmission at normal incidence, results that the outgoing wavepackets preserve the shape of the incoming wave. 

We consider a low-energy electron propagating in an infinite graphene sample, so that the system mimics a propagating massless Dirac particle, obeying $i\hbar\partial \Psi(\vec x, t)/\partial t = H\Psi(\vec x, t)$, for the Dirac Hamiltonian of graphene $H = v_F \vec{p}\cdot\vec\sigma$. The electron is described by a Gaussian wave packet, multiplied by a spinor that accounts for the probability distributions over the two sublattices of graphene (labeled A and B), and by a plane wave, which gives the particle a non-zero average momentum $k_0$:
\begin{eqnarray}
\Psi(\vec x,0) = N
\left(\begin{array}{c}
A\\
B
\end{array}\right) \exp \left[-\frac{(x-x_0)^2}{d_x^2}-\frac{(y-y_0)^2}{d_y^2} + i k_0 x\right],
\end{eqnarray}
where $N$ is a normalization factor, $(x_0,y_0)$ are the coordinates of the center of the Gaussian wave packet, $d_x$ ($d_y$) is its width in the $x$($y$)-direction, $k_0 = E_0/\hbar v_F$, and $E_0$ is the initial wave packet energy. The time evolution of such a wave packet is calculated by means of the split-operator technique, which is explained in details in Refs. [\onlinecite{Chaves0, Chaves1, Chaves2, Chaves3, Degani}]. However, in order to deal with a time-dependent potential $V(\vec{x},t)$, one must adapt the technique so that the time evolution operator in $\Psi(\vec x, t) = \hat U(0,t) \Psi(\vec x, 0)$ is written as
\begin{eqnarray}\label{eq.Operator}
\hat U(0,t) = e^{-\frac{i}{\hbar}\int_0^{t}H dt}.
\end{eqnarray}
The integral in the argument of the exponential in Eq. (\ref{eq.Operator}) can be separated into a sum of integrals for each time step, so that, in principle, one could make $U(0,t) = e^{-\frac{i}{\hbar}\int_{t-\Delta t}^{t}H dt}e^{-\frac{i}{\hbar}\int_{t-2\Delta t}^{t-\Delta t}H dt}...e^{-\frac{i}{\hbar}\int_{\Delta t}^{2\Delta t}H dt}e^{-\frac{i}{\hbar}\int_0^{\Delta t}H dt}$, which is then just an sequence of exponential operators applied recursively to an initial wave packet. However, if the Hamiltonian at two different time steps $t_1$ and $t_2$ does not commute, $[H(t_1),H(t_2)] \neq 0$, one has, in general, 
\begin{equation}
e^{-\frac{i}{\hbar}\int_{t_0}^{t_1}H dt -\frac{i}{\hbar}\int_{t_1}^{t_2}H dt} \neq e^{-\frac{i}{\hbar}\int_{t_0}^{t_1}H dt} e^{ -\frac{i}{\hbar}\int_{t_1}^{t_2}H dt}
\end{equation}
Nevertheless, the multiplication between exponentials in the right side of the equation can be seen as a first order approximation, according to the Suzuki-Trotter expansion of the exponential of a sum of non-commuting operators.\cite{Suzuki1, Suzuki2} The error in this approximation is proportional to $\Delta t^2$. Therefore, assuming a small $\Delta t$, applying the exponentials recursively to an initial wave packet provides, to a good approximation, an accurate description of the actual time-evolution of the system. Moreover, one can re-write
\begin{eqnarray}
\hat U(t,t+\Delta t) = e^{-\frac{i}{2\hbar}\int_t^{t+\Delta t}V dt}e^{-\frac{i}{\hbar} \hat T \Delta t}e^{-\frac{i}{2\hbar}\int_t^{t+\Delta t}V dt} + \mathcal{O}(\Delta t^3),
\end{eqnarray} 
where $\hat T$ is the kinetic energy operator. We further neglect the $\mathcal{O}(\Delta t^3)$ terms in this expansion by using a sufficiently small time step: since the argument of the exponential basically involves $\Delta t(T+V)/\hbar$, $T$ and $V$ are of the order of hundreds of meV and $\hbar$ is of the order of hundreds of meVfs, a fraction of fs (namely, $\Delta t = 0.1$ fs) is enough to give an accurate result. 

\begin{figure}[!b]
\centerline{\includegraphics[width = 0.8\linewidth]{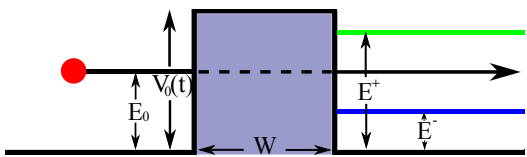}}
\caption{Sketch of the potential barrier (shaded area) considered in our calculations, which has width $W$ and a time-dependent height $V_0(t)$. The red circle represents the incoming electron, with energy $E_0$, which is able to go across the barrier, represented as a dashed line, even if $E_0 < V_0(t)$ for a given time $t$, due to Klein tunnelling. The outgoing electron leaves the barrier region either with higher ($E^+$) or lower ($E^-$) energy, depending on how the barrier height varies on time (see text).} \label{fig:Fig1}
\end{figure}

The wave packet starts at $x_0 = -300$ \AA\, and propagates through a potential barrier at $x = 0$, with width $W$, as illustrated by the shaded area in Fig. \ref{fig:Fig1}, where the barrier height is considered to vary linearly in time as $V_0(t) = \alpha t$ if $0 < x < W$, and $V = 0$ otherwise. As we explain in greater detail below, such a time-dependent potential can be generated \textit{e.g.} by a microwave field so that, for the typical spatial dimensions of the system, the period of oscillation of the potential is much smaller than the traversal time, thus retaining the linear dependence of the potential in time. In this case, the integral term in Eq. (\ref{eq.Operator}) becomes $\int_t^{t+\Delta t}V dt = \alpha(\Delta t^2 + 2t\Delta t)/2$ for $0 < x < W$.

After undergoing Klein tunnelling through a barrier, the propagating wave packet acquires a phase \cite{Savelev} given by
\begin{equation}
\phi = \frac{1}{\hbar v_F}{\int_{-\infty}^{\infty}V(x,t)dx}
\end{equation}
which, for the step barrier potential considered here yields $\phi = {W \alpha t}\big/{\hbar v_F}$. However, such a phase appears in the wave function as $\exp\left(-i\phi\right) = \exp\left({-\frac{i}{\hbar}\frac{\alpha W}{v_F}t}\right)$, which is equivalent to an extra energy factor $\exp\left({-\frac{i}{\hbar}E_x t}\right)$, where $E_x = \alpha W/v_F$. This suggests that after tunnelling through such a time varying potential, the electron would acquire an extra energy $E_x$. If $\alpha < 0$, the tunnelled electron loses energy and may even end up in a valence band state, provided the magnitude of $\alpha$ is high enough. This is indeed the case, as we will demonstrate. In all cases investigated here, time-dependence of the Hamiltonian did not affect the Klein tunnelling effect, and all transmission probabilities are found to be 1 (within the numerical precision of the calculation method).

\begin{figure}[!b]
\centerline{\includegraphics[width = 0.8\linewidth]{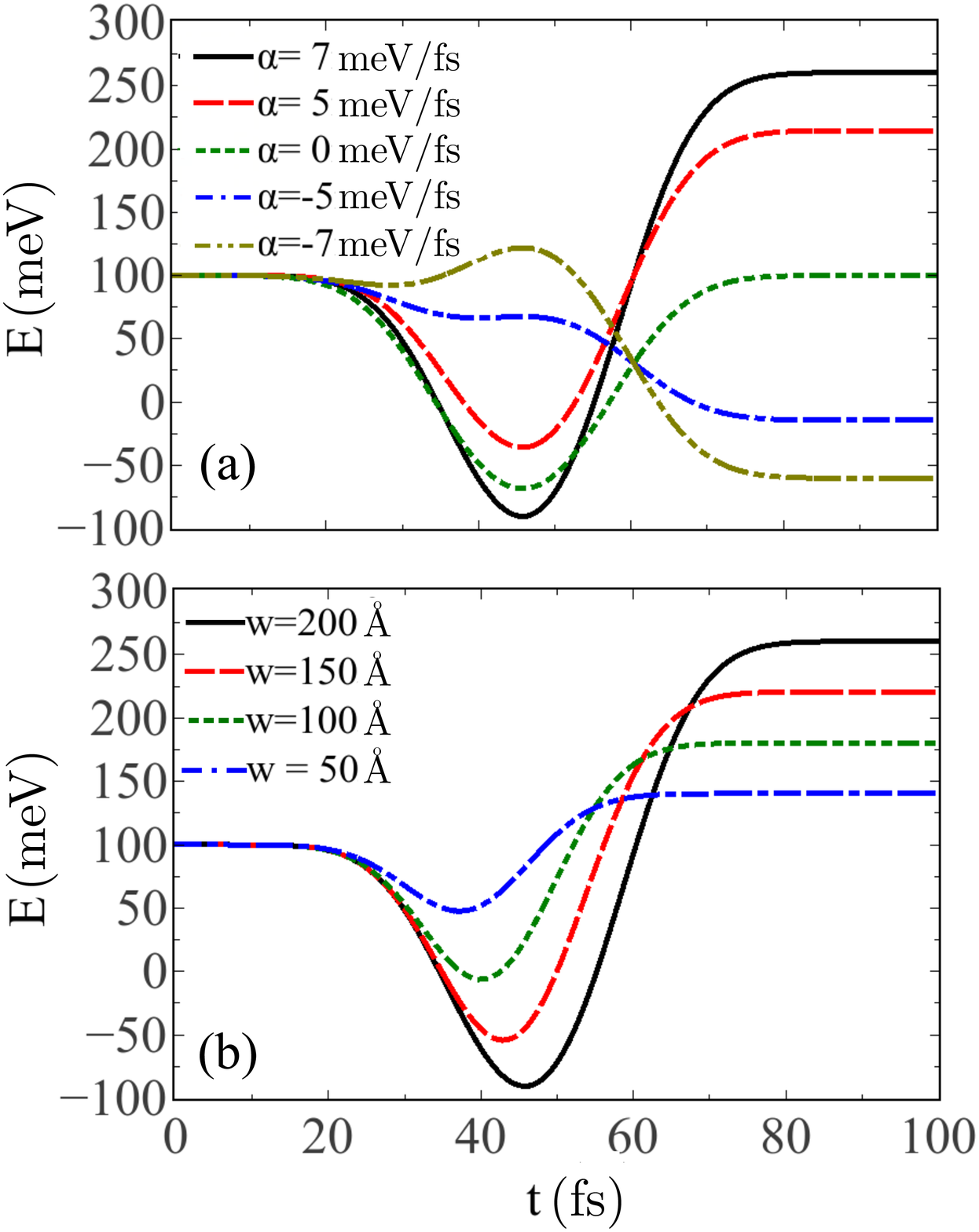}}
\caption{Kinetic energy of the wave packet as function of time for (a) a fixed width $W =$ 200\AA\, and different values of $\alpha$, and for (b) a fixed $\alpha = 7$ meV/fs and different values of width $W$.} \label{fig:Fig2}
\end{figure}

Let us first investigate a one-dimensional problem by considering $d_x = 100$ \AA\, and $d_y \rightarrow \infty$, which represents a wave front propagating in the $x-$direction. The initial energy of such a wave front is considered as $E_0 = 100$ meV. The average kinetic energy $E = \langle \Psi |T| \Psi \rangle$ is shown in Fig. \ref{fig:Fig2} as function of time for (a) a fixed width $W = $200\AA\, and several values of variation rate of the potential height $\alpha$, and for (b) a fixed $\alpha$ and several values of $W$. It is seen that the final energy $E$, after the wave packet leaves the barrier region, differs from the initial wave packet $E_0$ if $\alpha \neq 0$. In fact, the final energy is higher (lower) than $E_0$ when $\alpha > 0$ ($\alpha < 0$). That agrees with the argument that the energy shift is proportional to the rate of variation of the barrier height. Notice that in the $\alpha = -5$ and -7 meV/fs cases, the final energy is even negative, suggesting that the electron ended up in the valence band after traversing the time-dependent barrier. Notice that in all cases investigated here, we assume empty bands, so that all states are accessible by the electron and no Pauli blocking is involved. That would be the case, for example, of hot electrons in a system with a low Fermi level.

In the $\alpha \leq 0$ cases, we have considered an initial potential barrier of 200 meV at $t = 0$ fs. This explains why even for $\alpha = 0$ (i.e., for a constant 200 meV potential barrier), the wave packet loses kinetic energy while inside the barrier region: such energy reduction is related only to the addition of the average potential energy $\langle V \rangle$, which is non-zero only when the wave packet lies within the barrier. Nevertheless, the final energy does not depend on this initial potential barrier height, but rather only on $\alpha$, since it is a consequence of a different mechanism, namely, the equivalence between a Klein tunneling phase that depends linearly in time and an extra energy for the particle. Indeed, after leaving the barrier region in the $\alpha = 0$ case, the wave packet recovers its initial energy [see Fig. \ref{fig:Fig2}(a), green-dotted curve], no matter how high or low the barrier height is, as expected.

\begin{figure}[!t]
\centerline{\includegraphics[width = \linewidth]{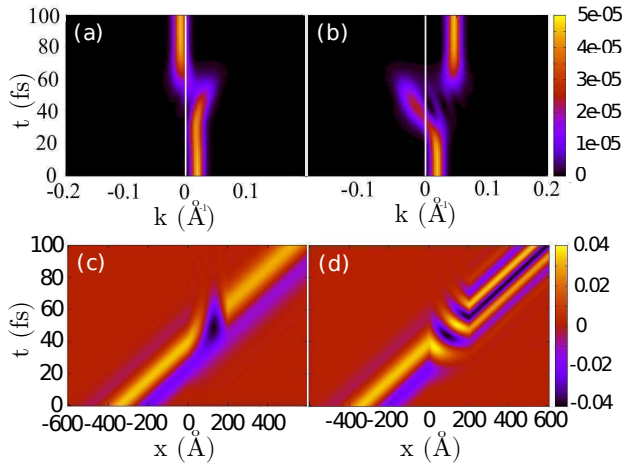}}
\caption{Contour plots of the time evolution of (a, b) the squared modulus of the wave packet in reciprocal space, as well as the (c, d) real part of the wave packet in the real space. The potential barrier lies within $0 \leq x \leq 200$ \AA\, and two values are considered for the linear dependence of its height in time: $\alpha$ = -7 meV/fs (a, c) and +7 meV/fs (b, d).} \label{fig:Fig3}
\end{figure}

Figure \ref{fig:Fig3} helps us to visualize the effect shown in Fig. \ref{fig:Fig2} by showing contour plots of the wave packet in reciprocal space (upper panels) as function of time. After leaving the barrier region, namely for $t \gtrapprox 60$ fs, the constant peak seen in the wave function at $k =0.0092$ \AA$^{-1}$ for $t = 0$ moves either (a) to the left, representing an energy decrease for $\alpha = -7$ meV/fs, even reaching the region of negative $k$, or (b) to the right, representing a energy enhancement for $\alpha = 7$ meV/fs. Despite the negative $\langle k \rangle$ in (a), we observe that in both cases, the \textit{group} velocity remains positive, as the wave packet keeps moving forward in real space. It is then interesting to see what happens to the \textit{phase} velocity in the $\langle k \rangle < 0$ case. Figures \ref{fig:Fig3}(c) and (d) show the contour plots of the real part of the wave function as function of time, for the same parameters as in Figs. \ref{fig:Fig3}(a) and (b), respectively. By the number of peaks, one can estimate the wave length, which is inversely proportional to $|k|$. After the barrier, the number of peaks remains essentially the same in (c) and increases in (d). The latter is consistent with the larger $\langle k \rangle$ (smaller wave length) observed for $t \gtrapprox 60$ fs in Fig.\ref{fig:Fig3}(b), whereas the former shows an outgoing wave packet whose wave vector $k$ has changed in sign, but not significantly in magnitude. We observe that phase velocity is always positive after the barrier (the position of each peak increases with time), which confirms that the outgoing wave packet in Fig. \ref{fig:Fig3}(a) behaves as a hole with $E < 0$, since the phase velocity is $v_p = E/\hbar k > 0$ and $\langle k \rangle$ is shown to be negative after the barrier in this case.

\begin{figure}[!b]
\centerline{\includegraphics[width = 0.8\linewidth]{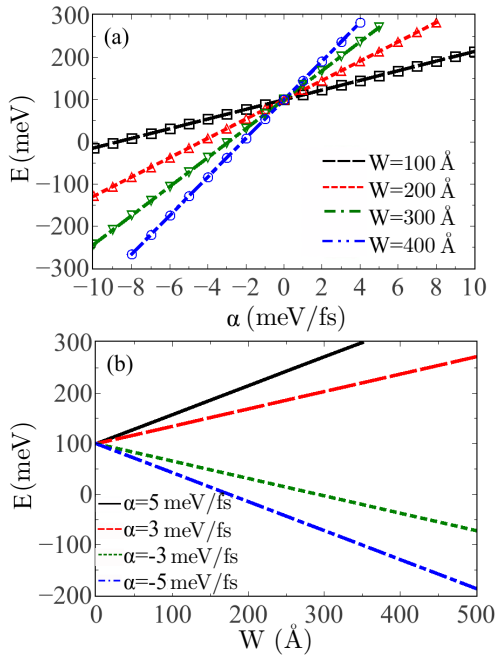}}
\caption{Final energy, after the wave packet leaves the potential barrier, (a) as function of the variation rate of the potential height $\alpha$, considering different values of barrier width $W$, and (b) as function of $W$, for different values of $\alpha$. Symbols in (a) are the numerically obtained results, whereas curves are an analytical estimate.} \label{fig:Fig4}
\end{figure}

Our previous analytical calculations show that the extra energy $E_x$ provided by the time-dependent potential barrier depends linearly both on $\alpha$ and $W$. This is confirmed by the results in Fig. \ref{fig:Fig4}, which show the final energy $E_{final}$ as function of these quantities. In Fig. \ref{fig:Fig4}(a), the numerically obtained final energy $E$ (symbols) is compared to the one predicted by the Klein tunnelling phase for a time dependent barrier (lines), as explained above, where very good agreement is observed. Figure \ref{fig:Fig4}(b) confirms that this extra energy depends linearly on $W$ and demonstrates that electronic wave packets can become hole-like for strongly negative $\alpha$ - considering an initial wave packet energy $E = 100$ meV and $\alpha = -5$ meV/fs, e.g., the final energy is $E < 0$ for $W \gtrsim 150$ nm.

\begin{figure}[!t]
\centerline{\includegraphics[width = \linewidth]{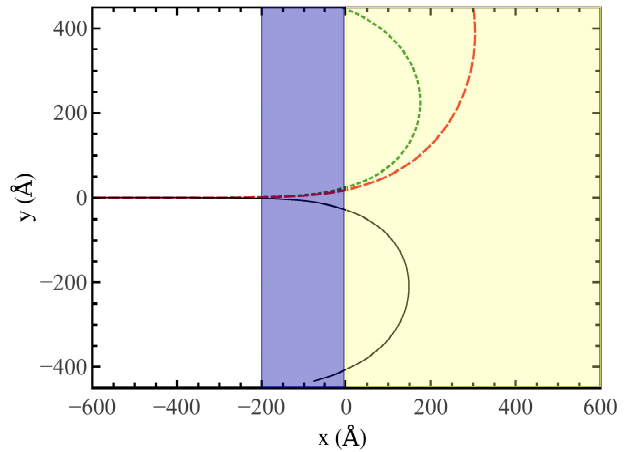}}
\caption{Trajectories of a circular wave packet $d_x = d_y = 200$ \AA\, tunnelling through a barrier at $-200$ \AA\, $\leq x < 0$ (purple, dark-shaded area) and reaching a magnetic step barrier for $x \geq 0$ (yellow, light-shaded area). The wave packet starts with initial energy $E_0 = 100$ meV, and the barrier height varies with rates $\alpha = -8.2$ meV/fs (black solid) and $\alpha = +4.1$ meV/fs (red dashed). The result for $\alpha = 0$ (green dotted) is shown for comparison.} \label{fig:Fig5}
\end{figure}

Further indication of the conduction-to-valence band states conversion in the case of high negative $\alpha$ can be gained by analysing the behavior of the outgoing wave packet in the presence of an external magnetic field. For instance, if the outgoing packet represents a valence band state, which behaves as positively charged quasi-particle (similar to a hole in semiconductors or a positron in high-energy systems), its behavior under such field should be opposite to that of an electron. In order to check that, we carried out the following simulation: a circular wave packet $d_x = d_y = 200$ \AA\, with energy $E_0 = 100$ meV starts at $x_0 = -600$ \AA\,, reaches a $W$ = 200 \AA\, barrier, placed between $x = -200$ \AA\, and $x = 0$, and enters a $B = 5$ T magnetic barrier for $x > 0$. The trajectories of the center of mass of such wave packet, calculated by $(\langle x \rangle, \langle y \rangle)$, resulting from such simulation, are shown in Fig. \ref{fig:Fig5}, considering $\alpha = 0$ (green dotted), 4.1 (red dashed) and -8.2 meV/fs (black solid). By the fact that the trajectory for $\alpha = 4.1$ meV/fs exhibits larger radius as compared to the one for $\alpha = 0$, one already infers that the time-dependent barrier has boosted the electron energy, since the radii of the circular trajectories coming from Lorentz force are directly proportional to the particle energy $E$ as $R = E\big/e v B$. \cite{Chaves1} In fact, the trajectory for the $\alpha = 0$ case resembles a semi-circle with a diameter $2R \approx 450$ \AA\, inside the magnetic barrier, whereas for $\alpha = 4.1$ meV/fs, which leads to an extra energy $E_x \approx$ 100 meV and, consequently, doubles the electron energy, the resulting circular trajectory has twice larger diameter. Moreover, after passing through the $\alpha = -8.2$ meV/fs barrier, the wave packet trajectory is curved to the opposite direction, which is expected for a positively charged particle, but with a similar radius as the one observed for $\alpha = 0$ (green dotted). In fact, $\alpha = -8.2$ meV/fs leads to an energy loss of $\approx 200$ meV, which makes the wave packet, which initially has $E_0 =$ 100 meV, end up with $E \approx -100$ meV, i.e. as a hole. As the final energies for $\alpha = 0$ and $\alpha = -8.2$ meV/fs differ practically only by a negative sign, it is then expected that their circular trajectories inside the magnetic barrier exhibit almost the same radii, but in opposite directions.

It is worthy to point out the difference between the valence band quasi-particles observed here and the usual definition of holes in semiconductor physics. The latter is rather a collective effect of valence band electrons: if one of the valence band states is unoccupied, the dynamics of the remaining electrons in real and reciprocal spaces can be effectively described by that of a positively charged particle, which even interacts via Coulomb potential with conduction band electrons, forming excitons, trions, etc. Holes in semiconductors can be also seen as an analog of a positron in the context of high-energy physics. \cite{Tunneling1} In our study, the valence band is empty, so, we cannot have a hole \textit{strictu sensu}. What is observed is rather regarded as a valence band electron, whose charge is still negative, so that charge conservation is respected. However, interestingly enough, such state turns out to behave as a positively charged particle (just like holes and positrons), say, in the presence of electromagnetic fields, due to its negative effective mass.   

It is important to address the issue of the feasibility of experimental detection of such an effect. The time dependent barrier could be provided by a time dependent electric field perpendicular to the layer and applied just in a finite region, so that electrons in this region (barrier) would have higher energy. As mentioned above, such a time dependent electric field can be obtained from an electromagnetic wave. Linear dependence of the potential, however, would only be obtained by approximation of the sinusoidal form of the electromagnetic wave as $sin(\omega t) \approx \omega t$, which is only valid for $t \ll 1/\omega$. Therefore, the time for the electron to cross the whole barrier $t = W/v_F$ must be much smaller than the period of the electromagnetic wave in order to produce a barrier with approximately linear time dependence, so $\omega \ll v_F/W$. As an example, a barrier with width $W = 1000$ \AA\,, for instance, would require $\omega \ll 8.2$ THz, i.e. in the higher frequency limit of the GHz (microwaves) range. On the other hand, such electromagnetic wave would produce a potential $V_0(t) = -\Phi_0 \sin(\omega t) \approx \Phi_0 \omega t$, where $\Phi_0 = Fd$, $d$ is the distance from the source to the graphene sample and $F$ is the electric field intensity, therefore $\Phi_0\omega$ plays the role of $\alpha$ in our theory. If one considers specifically the conduction-to-valence band states transition, the amount of energy $E_x$ reduced by the time dependent barrier in this case must be larger than the initial electron energy, $E < E_x$. After some algebraic manipulations, one finds $\omega = E_x v_F \big/\Phi_0 W \ll t$, which yields $\Phi_0 \gg E$. In other words, the conduction-to-valence band states transition can be observed provided the electromagnetic wave frequency is in the GHz range (to guarantee linearity of the potential in time) and its intensity is high enough. 

In summary, we investigated the Klein tunnelling of an electronic wave packet through a potential barrier whose height varies linearly in time, in an infinite monolayer graphene sample. Our results demonstrate that for this specific form of time dependence of the potential barrier, the phase acquired by the electron after the tunnelling has an important meaning, as it can be seen as an extra energy for this charge carrier. If this extra energy is negative, an incident electron, initially in the conduction band, can be converted into a valence band state after tunnelling, which behaves similarly to a positively charged quasi-particle. Such transition is theoretically verified both by analysing the sign of the tunnelled wave packets energy and the trajectory of the center of mass of such wave in the presence of an external magnetic field. Results predicted here are possible to be experimentally observed e.g. by investigating the scattering of hot electrons by an oscillating potential barrier, whose height is modulated by a microwave \cite{Kowenhoven} focused on a finite region of the graphene sample.

\acknowledgements Discussions with F. M. Peeters are gratefully acknowledged. This work was financially supported by CNPq, under the PRONEX/FUNCAP grants and CAPES.


\begin{references}

\bibitem{textbook} C. Cohen-Tannoudji, B. Diu, and F. Laloe, Quantum Mechanics, 1st ed., Vol. 1, Chap. I, compl. H-I (WileyInterscience, New York, 1978).

\bibitem{Tunneling1} M. I. Katsnelson, K. S. Novoselov, and A. K. Geim, Nature Phys. \textbf{2} 620 (2006). 

\bibitem{Tunneling2} A. F. Young and P. Kim, Nature Phys. \textbf{5}, 222 (2009).

\bibitem{Tunneling3} P. E. Allain and J. N. Fuchs, Eur. Phys. J. B \textbf{83}, 301 (2011).

\bibitem{Chaves2} J. M. Pereira, F. M. Peeters, A. Chaves and G. A. Farias, Semicond. Sci. Technol. \textbf{25}, 033002 (2010).

\bibitem{Logemann} R. Logemann, K. J. A. Reijnders, T. Tudorovskiy, M. I. Katsnelson, and Shengjun Yuan, Phys. Rev. B \textbf{91}, 045420 (2015).

\bibitem{AKGeim} K. S. Novoselov, A. K. Geim, S. V. Morozov, D. Jiang, Y. Zhang, S. V. Dubonos, I. V. Grigorieva, and A. A. Firsov, Science \textbf{306} 666 (2004).

\bibitem{CastroNetoReview} A. H. Castro Neto, F. Guinea, N. M. R. Peres, K. S. Novoselov and A. K. Geim, Rev. Mod. Phys. \textbf{81}, 109 (2009).

\bibitem{Misha}
M. I. Katsnelson, {\it Graphene: Carbon in Two Dimensions} (Cambridge University Press, 2012).

\bibitem{time1} D. L. Haavig, and R. Reifenberger, Phys. Rev. B {\bf 26}, 6408 (1982).

\bibitem{time2} M. L. Chiofalo, M. Artoni, and G. C. La Rocca, New J. Phys. {\bf 5}, 78 (2003).

\bibitem{time3} R. M. Dimeo, Am. J. Phys. {\bf 82}, 142 (2014).

\bibitem{time4} G. Sulyok, J. Summhammer, and H. Rauch, Phys. Rev. A \textbf{86}, 012124 (2012).

\bibitem{time5} Shi-Jun Liang, S. Sun, and L.K. Ang, Carbon \textbf{61}, 294 (2013). 

\bibitem{Chaves0} A. Chaves, G. A. Farias, F. M. Peeters, and R. Ferreira, Comm. Comp. Phys. \textbf{17}, 850 (2015).

\bibitem{Chaves1} Kh. Yu. Rakhimov, A. Chaves, G. A. Farias and F. M. Peeters, J. Phys.: Condens. Matter \textbf{23}, 275801 (2011).

\bibitem{Chaves3} A. Chaves, L. Covaci, Kh. Yu. Rakhimov, G. A. Farias, and F. M. Peeters, Phys. Rev. B \textbf{82}, 205430 (2010).

\bibitem{Degani} M. H. Degani and M. Z. Maialle, J. Comput. Theor. Nanosci. 7, 454 (2010).

\bibitem{Suzuki1} M. Suzuki, Phys. Lett. A \textbf{387}, 165 (1992).
 
\bibitem{Suzuki2} M. Suzuki, J. Math. Phys. \textbf{26}, 601 (1985).

\bibitem{Savelev} S. E. Savelev, W. H\"ausler, and Peter H\"anggi, Phys. Rev. Lett. \textbf{109}, 226602 (2012).

\bibitem{Kowenhoven} L. P. Kouwenhoven, S. Jauhar, K. McCormick, D. Dixon, P. L. McEuen, Yu. V. Nazarov, N. C. van der Vaart, and C. T. Foxon, Phys. Rev. B \textbf{50}, 2019(R) (1994).


\end{references}
\end{document}